\documentclass[a4]{PoS}
\usepackage{enumitem}
\usepackage{url}
\usepackage{tikz}
\usepackage{rotating}
\usepackage{booktabs}
\usepackage[percent]{overpic}
\usepackage{xcolor}

\newcommand{\jpsi}{\ensuremath{J\!/\!\psi}}
\newcommand{\gev}{\hbox{ GeV}}
\newcommand{\mev}{\hbox{ MeV}}
\def\abs#1{\left| #1\right|}
\newcommand{\alphas}{\ensuremath{\alpha_{\mathrm{s}}}}
\newcommand{\ps}{\hbox{ ps}}
\newcommand{\cfrac}[2]{\scriptstyle \frac{#1}{#2}}

\newcommand{\arxiv}[1]{\href{https://arxiv.org/abs/#1}{arXiv:#1}}
\newcommand{\doi}[2]{\href{https://doi.org/#1}{#2}}
\newcommand{\zenodo}[1]{\href{https://doi.org/10.5281/zenodo.#1}{zenodo.#1}}
\newcommand{\xw}{\ensuremath{\sin^2\!\theta_{\mathrm{W}}}}
\def\half{{\scriptstyle \frac{1}{2}}}
\newcommand{\spec}[4]{\ensuremath{#1^{#2}\!{#3}_{#4}}}
\newcommand{\fm}{\hbox{ fm}}
\newcommand{\fb}{\ensuremath{\hbox{ fb}}}
\newcommand{\tev}{\ensuremath{\hbox{ TeV}}}

\definecolor{Dgreen}{RGB}{0,96,0}

\title{Beauty at High $\left\{\begin{array}{l}\hbox{\textsf{\textbf{Precision}}}\\[2pt] \hbox{\textsf{\textbf{Sensitivity}}}\end{array}\right.$}

\ShortTitle{Beauty 2019 Opening}

\author{\speaker{Chris Quigg}\\
        Theoretical Physics Department, Fermi National Accelerator Laboratory \\ P.O. Box 500, Batavia, Illinois 60510 USA\\
        E-mail: \email{quigg@fnal.gov}\qquad\href{https://orcid.org/0000-0002-2728-2445}{\tt ORCID: 0000-0002-2728-2445}}
\abstract{Opening lecture at Beauty2019.\hfill \textsf{FERMILAB--CONF--20-071--T}

\vspace*{12pt}
Origins of contemporary $B$-physics. Mesons with beauty and charm. Stable tetraquarks? Flavor and the problem of identity. Top matters. Electroweak symmetry breaking and the Higgs sector. Future instruments.\hfill Slides available at  \href{https://doi.org/10.5281/zenodo.3468909}{zenodo.3468909}.}

\FullConference{18th International Conference on B-Physics at Frontier Machines - Beauty2019 -\\
		29 September / 4 October, 2019\\
		Ljubljana, Slovenia}

\begin{document}

\section{Introduction}
As we open Beauty2019, I note with pleasure the large number of young scientists among the participants. Since many of you were not yet living when $B$ physics was born, I want to begin with a short review of our Origin Story.  I will next  touch on two topics in hadron spectroscopy that have been particularly interesting to me recently: next steps in the investigation of the $B_c$ spectrum and the likely existence of doubly heavy tetraquarks that are stable, or nearly stable, against strong decay. Then I will speak more generally to the future of our subject, posing  questions about flavor physics, the top quark, and electroweak symmetry breaking and the Higgs sector. In anticipation of the European Strategy Update for Particle Physics~\cite{EuroStrat}, I will close by inviting you to consider the relative merits of future accelerator projects.

\section{Origin Story}
\begin{figure}[h]
\centerline{\includegraphics[height=0.2\textheight]{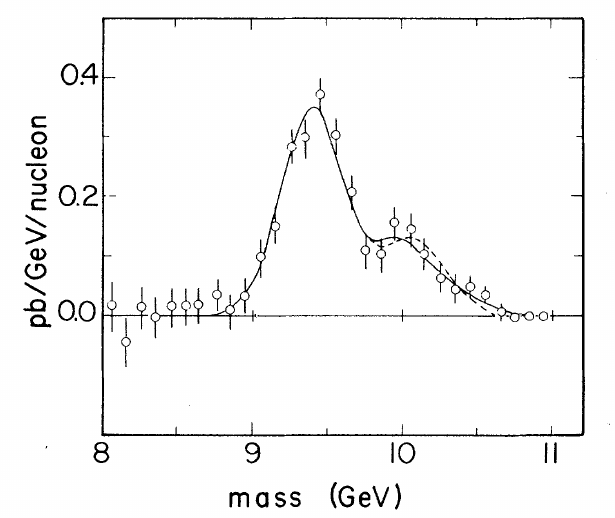}\quad
\raisebox{56pt}{\begin{tabular}{|ccc|}
\hline
E288 & $M(\Upsilon^\prime) - M(\Upsilon)$ & $M(\Upsilon^{\prime\prime}) - M(\Upsilon^\prime)$ \\
\hline
Two-level fit & $650 \pm 30\mev$ & \\
Three-level fit & $610 \pm 40\mev$ & $1000 \pm 120\mev$ \\
$M(\psi^\prime) - M(\jpsi)$ & $\approx 590\mev$ & \\ \hline
\end{tabular}}}
\caption{Left panel: The dimuon mass distribution in the reaction $pN \to \mu^+\mu^-+X$, showing at least two $\Upsilon$ resonance peaks (from Ref.~\cite{Innes:1977ae}). The solid curve is the three-peak fit; the dashed curve is the two-peak fit. Right panel: Fits  suggested an unresolved third peak.}
\label{fig:origin}
\end{figure}

The first experimental evidence for the existence of the fifth quark came in the summer of 1977, with the discovery of a strong enhancement at $9.5\gev$ in the mass spectrum of dimuons produced in collisions of 400-GeV protons with Cu or Pt targets~\cite{Herb:1977ek} at Fermilab. Later that year, a threefold increase in statistics made it possible to resolve at least two peaks consistent in width with experimental resolution~\cite{Innes:1977ae}. The excess of the data over a fit to the continuum fit is shown in Figure~\ref{fig:origin}. 
The resonances were designated  $\Upsilon$ and tentatively identified as bound states of a new heavy quark and antiquark, by analogy with the charmonium $(c\bar{c})$ family. It was noteworthy that the spacing between the (apparent) ground state and first excited state was very similar to the mass difference between $\psi^\prime(3686)$ and $\jpsi(3097)$.

Experiment 288, as it was known, was proposed before the November 1974 Revolution~\cite{e288}. It promised to search for structures in the dilepton spectrum, ``publish these and become famous.'' The subsequent discoveries of the charmonium resonances~\cite{Aubert:1974js,Augustin:1974xw} and the $\tau$ lepton~\cite{Perl:1975bf} precipitated a wave of dilepton experiments at Fermilab, which the CERN \emph{Courier} characterized as dileptomania~\cite{dilepto}. Although Makoto Kobayashi and Toshihide Maskawa's insight~\cite{Kobayashi:1973fv} that three generations of quarks could enable \textsf{CP} violation through the complex phase in the $3 \times 3$ quark-mixing matrix had been published, the inevitability of a third generation had not yet taken hold in the community. 

The evidence for three narrow peaks was in accord with what Eichten \& Gottfried~\cite{Eichten:1976jk} had anticipated within a Coulomb $+$ linear potential model, in their preparations for the Cornell Electron Storage Ring Proposal in November 1976. They calculated quarkonium spectra over a range of quark masses $m_Q \gtrsim m_c$. At the 5-GeV nominal beam energy of CESR, they foresaw three narrow levels, as observed, but predicted a level spacing $E(2S)-E(1S) \approx 420\mev$.
It soon came to light that for a very general class of potentials, the number of narrow $^3$S$_1$ $Q\bar{Q}$ levels that
lie below the threshold for Zweig-allowed decay grows as
$N = a \sqrt{m_Q/m_c}$~\cite{Quigg:1977xd}. Since $N=2$ for charmonium, it is a general result that three or perhaps 4 narrow $\Upsilon$ levels should be seen, depending on the ratio of quark masses. Combining information from the \jpsi\ and $\Upsilon$ families, we would come to learn much more about the interquark potential than we could from either family alone.

Why choose $m_Q = 5\gev$? Fermilab experiment E1A had reported an excess of events at high values of the inelasticity parameter $y = (E_\nu - E_\mu)/E_\nu$  in the reaction
$\bar{\nu}_\mu N \to \mu^+ + \mathrm{anything}$~\cite{Benvenuti:1976ad}. The excess was dubbed the high-$y$ anomaly; for a left-handed charged current interaction, we would expect the behavior $d\sigma(\bar{\nu}q)/dy \propto (1-y)^2$ to characterize antineutrino scattering on a target made (mostly) of quarks, in contrast to the $d\sigma(\nu q)/dy \propto 1$ behavior expected for neutrinos on quarks. The excess events could be explained by a right-handed $u \to b$ transition with $m_b \approx 4 \hbox{ -- } 5\gev$~\cite{Barnett:1976kh}. That was not to be. In an interesting dramatic twist, Leon Lederman's announcement of the $\Upsilon$ discovery at the 1977 European Physical Society Meeting in Budapest was immediately preceded by Jack Steinberger's report that the CDHS experiment had ruled out the high-$y$ anomaly~\cite{Holder:1977en}!

\begin{figure}[h]
\hspace*{3em}{\begin{overpic}[height=0.35\textheight]{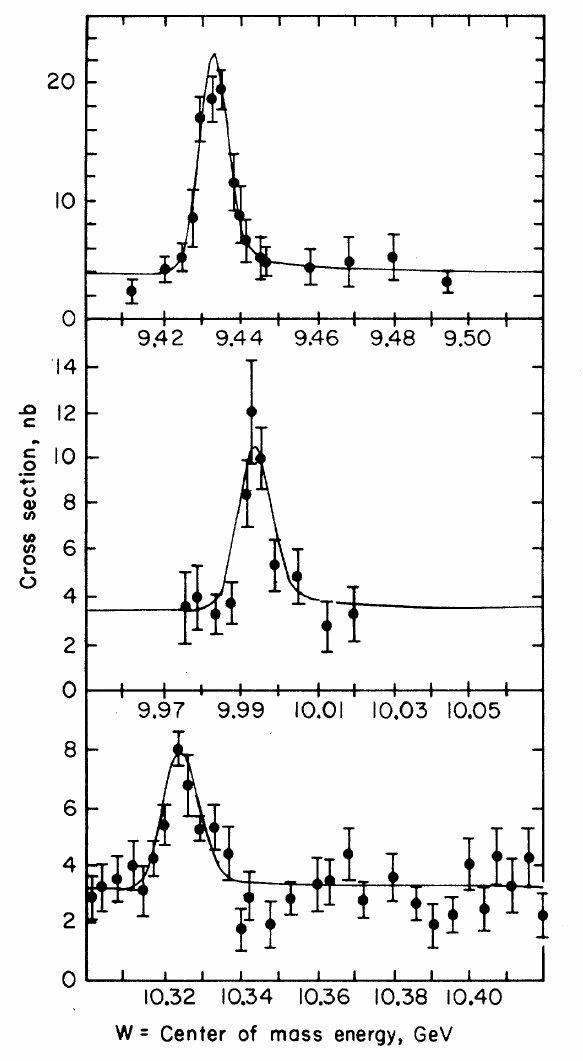}
 \put (30,90) {\Large \textsf{CLEO}}
\end{overpic}}
\hspace*{24pt}
\raisebox{56pt}{{\begin{overpic}[width=0.5\textwidth]{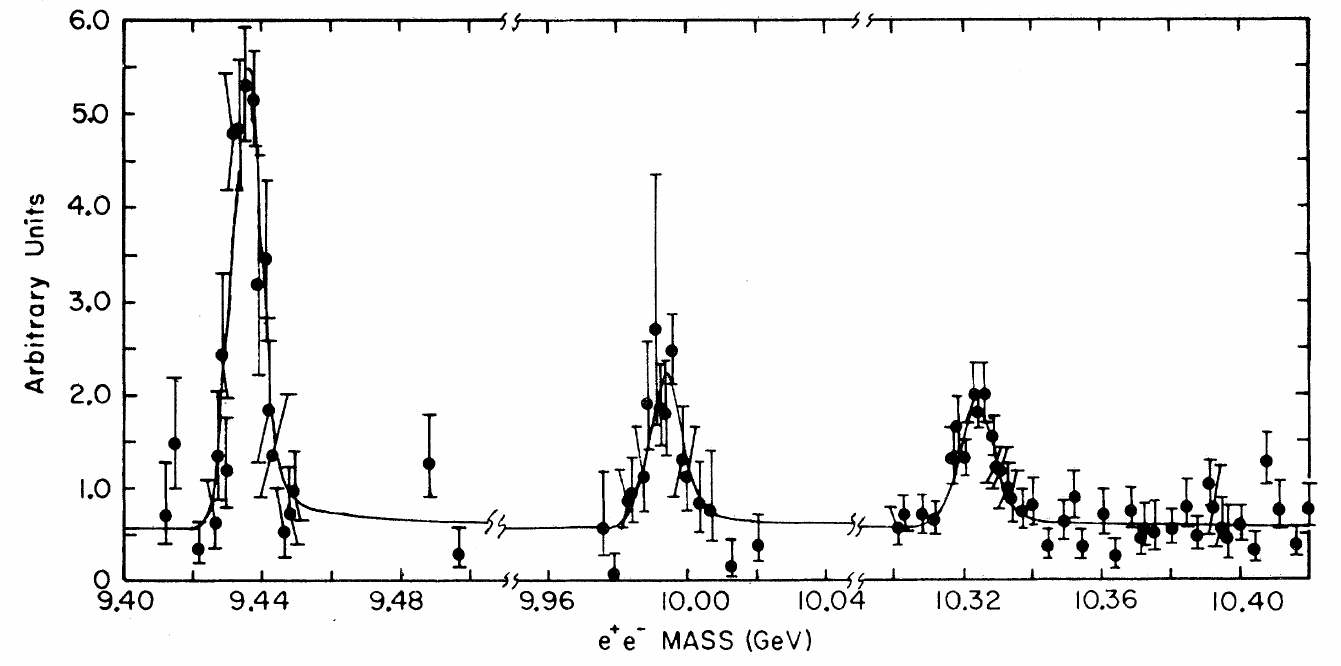}
\put(65,40){\Large \textsf{CUSB}}
\end{overpic}}}
\caption{Observation of three narrow $\Upsilon$ levels by the CLEO and CUSB experiments at CESR, Ref.~\cite{Andrews:1980ha}. \label{fig:ups3}}
\end{figure}

Despite the vanishing of a 5-GeV right-handed $b$ quark, both CESR and the DORIS storage ring at DESY had plenty to study, thanks to the discovery of the $\Upsilon$ family. A year after the discovery of $\Upsilon$, measurements of the $\Upsilon(1S), \Upsilon(2S)$ leptonic widths at DORIS pinned down the charge of the new quark as $Q_b = -\cfrac{1}{3}$~\cite{Jackson:1978he}. Then, over the 1979--1980 end-of-year holidays, two experiments at the Cornell Electron Storage Ring announced that they had  resolved three narrow $\Upsilon$ states~\cite{Andrews:1980ha}, confirming the suspicion raised by the discovery data from Fermilab (see Figure~\ref{fig:ups3}). 

In 1981, studies by the CLEO experiment~\cite{Bebek:1980zd} at the $\Upsilon(4S)$ resonance launched $B$ physics as we know it today. They observed an enhancement  in the inclusive cross section for directly produced single electrons on the resonance. This discovery was evidence for a new weakly decaying particle, the $B$
 meson, with a semileptonic branching ratio $\mathcal{B}(B \to X e \nu) \approx13\%$.
 
\begin{figure}[h]
\centering
\begin{overpic}[height=0.302\textheight]{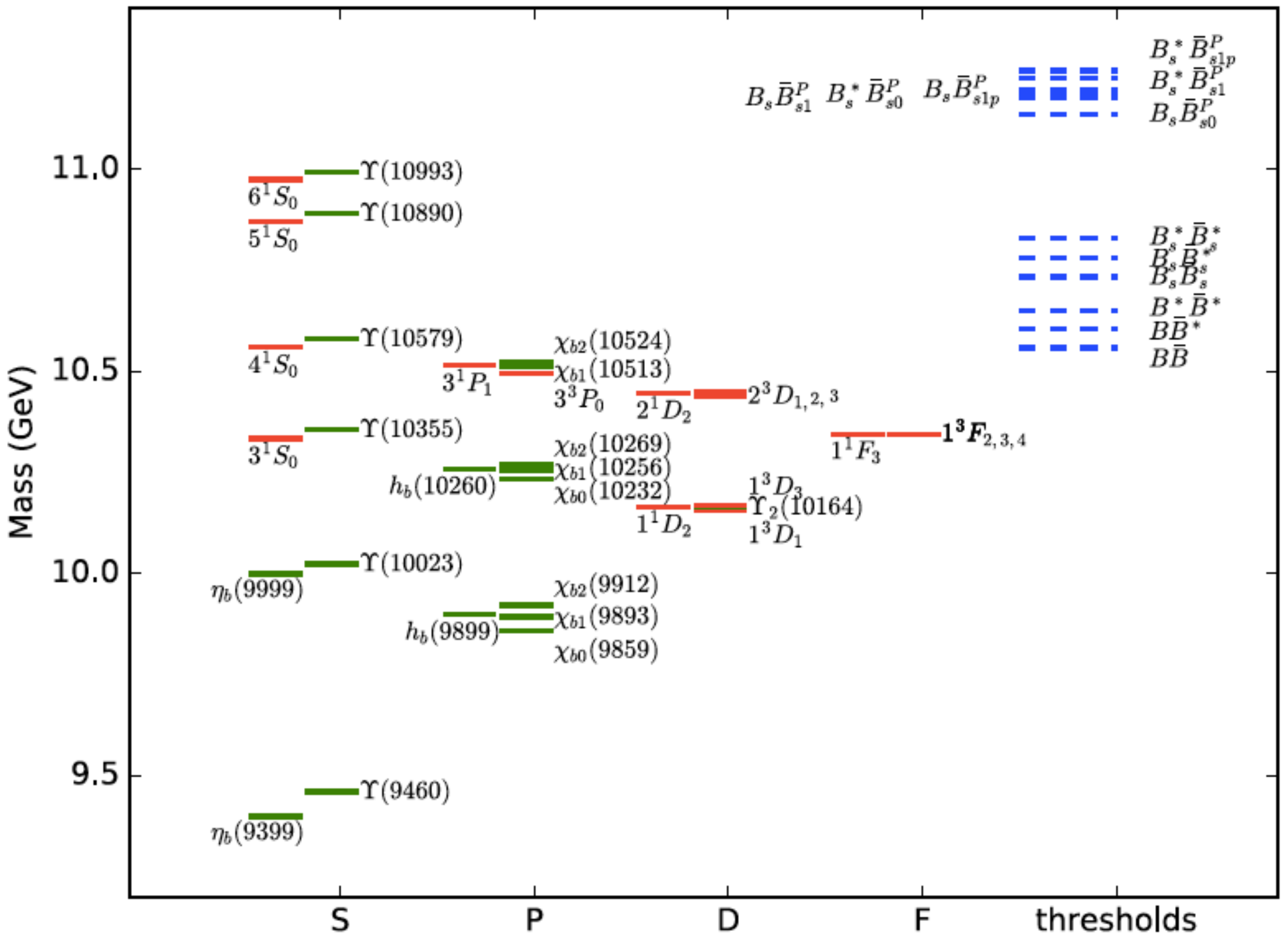}\put(75,20){\small\textcolor{Dgreen}{\textsf{Observed}}}\put(75,15){\small\textcolor{red}{\textsf{Predicted}}}
\end{overpic}      \caption{Predicted and observed levels of the $(b\bar{b})$ quarkonium family~\cite{EJEpc}. Two-meson flavor thresholds are shown as dashed lines to the right of the figure.}
       \label{fig:upsilons}
     \end{figure}

 Although the Particle Data Group's summary tables for $(b\bar{b})$ mesons run to seventeen pages,  a rich spectrum of levels is still to be explored---fourteen ordinary $(b\bar{b})$ states below flavor threshold, as indicated by the display  in Figure~\ref{fig:upsilons}.

\begin{figure}
\centerline{\begin{overpic}[height=0.308\textheight]{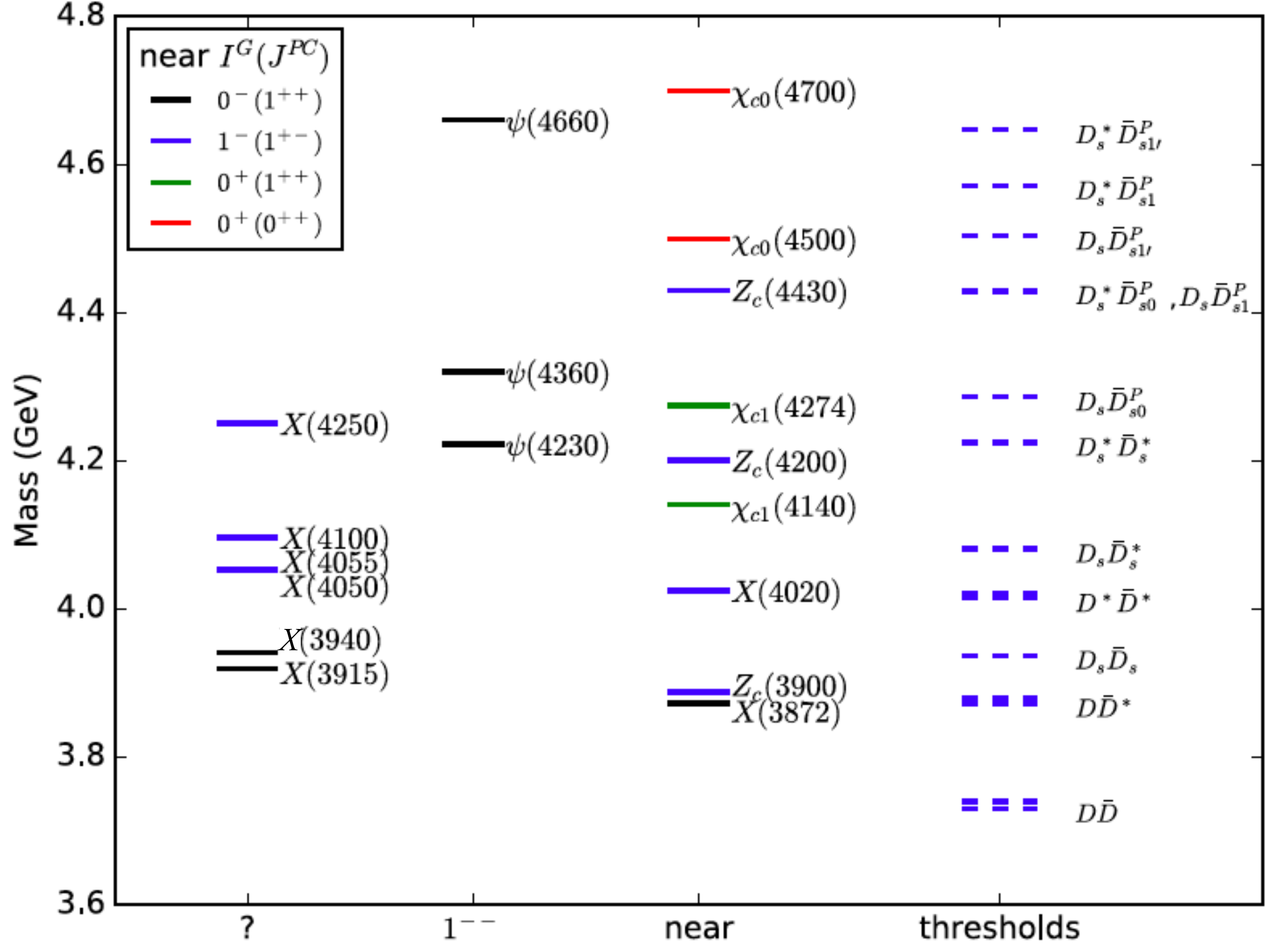}
\end{overpic}}
\caption{Some states associated with charmonium that may entail body plans beyond $(c\bar{c})$~\cite{EJEpc}. Neutral states not identified as $J^{PC} = 1^{--}$ are labeled by $X$, charged states  by $Z$. \label{fig:weirdos}}
\end{figure}

And that is sure to be far from the whole story. We have encountered many states associated with charmonium  that seem not to be pure $(c\bar{c})$ configurations~\cite{Olsen:2017bmm}, some  indicated in Figure~\ref{fig:weirdos}.

Beginning with the famous $X(3872)$, now designated $\chi_{c1}(3872)$ without prejudice to its composition, these above-flavor-threshold states are mostly narrow, and are seen in hadronic transitions or decays. Many lie in close proximity to two-meson thresholds, which may influence their properties, along with communication with open-flavor channels that must be considered for all states near or above threshold. Bearing in mind that the novel states are likely to be quantum-mechanical superpositions of several components, the response to ``What are they?'' may contain components of $(c\bar{c})$ quarkonium and of new body plans, including quarkonium hybrids $(q\bar{q}g)$ and $(qq\bar{q}\bar{q})$ states. In the latter category, we may include dimeson ``molecules,'' tetraquark mesons, diquarkonium, and hadroquarkonium. When can we find $(b\bar{b})$ analogues in similar profusion?

Prescient papers in 1980--1981 by Carter \& Sanda~\cite{Carter:1980hr} and Bigi \& Sanda~\cite{Bigi:1981qs} emphasized that
\textsf{CP} violation might be large and observable in $B$ decays.

\begin{figure}[h!]
\centerline{\includegraphics[height=0.27\textheight]{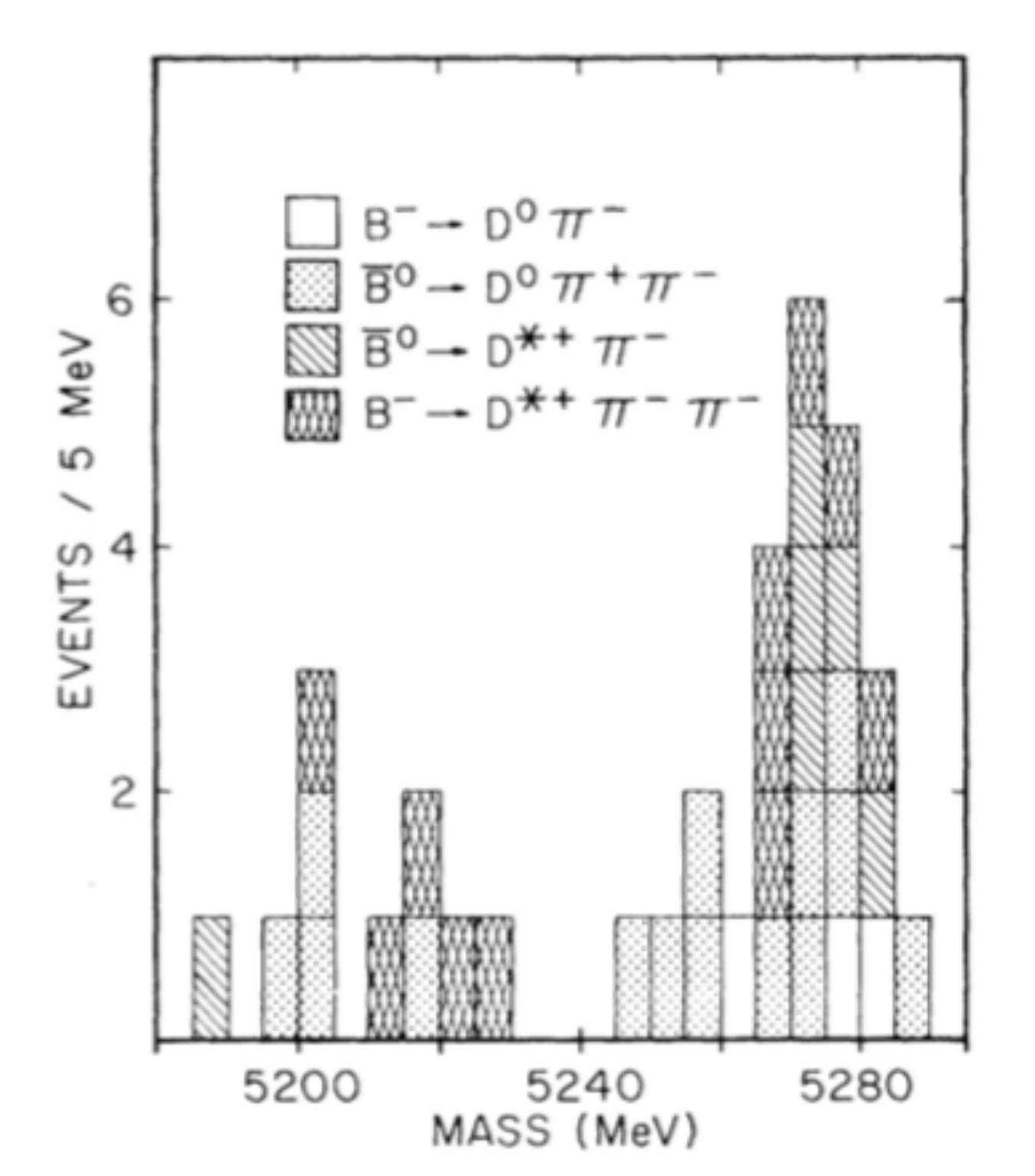}}
\caption{Mass distribution of $B$-meson candidates from the CLEO experiment, Ref.~\cite{Behrends:1983er}. The labels imply charge-conjugate final states as well.\label{fig:Brecon}}
\end{figure}

$B$ mesons were first reconstructed by the  CLEO Collaboration in 1983 in final states containing $D^0$  or $D^{*\pm}$ and one or two charged pions~\cite{Behrends:1983er}. I call your attention to the  low statistics shown in Figure~\ref{fig:Brecon}; what a contrast to the enormous samples available today! CLEO estimated the charged-$B$  mass  as $5270.8 \pm 2.3 \pm 2.0\mev$ and the neutral-$B$  mass as $5274.2 \pm 1.9 \pm 2.0\mev$, a few MeV lower than the current world averages, $M(B^\pm) = 5279.33 \pm 0.13\mev$ and $M(B^0) = 5279.64 \pm 0.13\mev$~\cite{Tanabashi:2018oca}.
According to the Particle Data Group, the quantum numbers $I, J, P$ of $B^\pm$ and $B^0$ still require confirmation.

\begin{figure}
\begin{minipage}[b]{0.29\textwidth}
\begin{center}
Charm lifetimes [fs]\\
$D^+: 1040 \pm 7$\\
$D^0: 410.1 \pm 1.5$\\
$D_s: 504 \pm 4$\\
$\Lambda_c: 200 \pm 6$\\
$\Xi_c^+: 442 \pm 26$\\
$\Xi_c^0: 112^{+13}_{-10}$\\
$\Omega_c: 268 \pm 26$

\vspace*{64pt}
\end{center}
\end{minipage}
\begin{minipage}[t]{0.40\textwidth}
\centerline{\includegraphics[width=\textwidth]{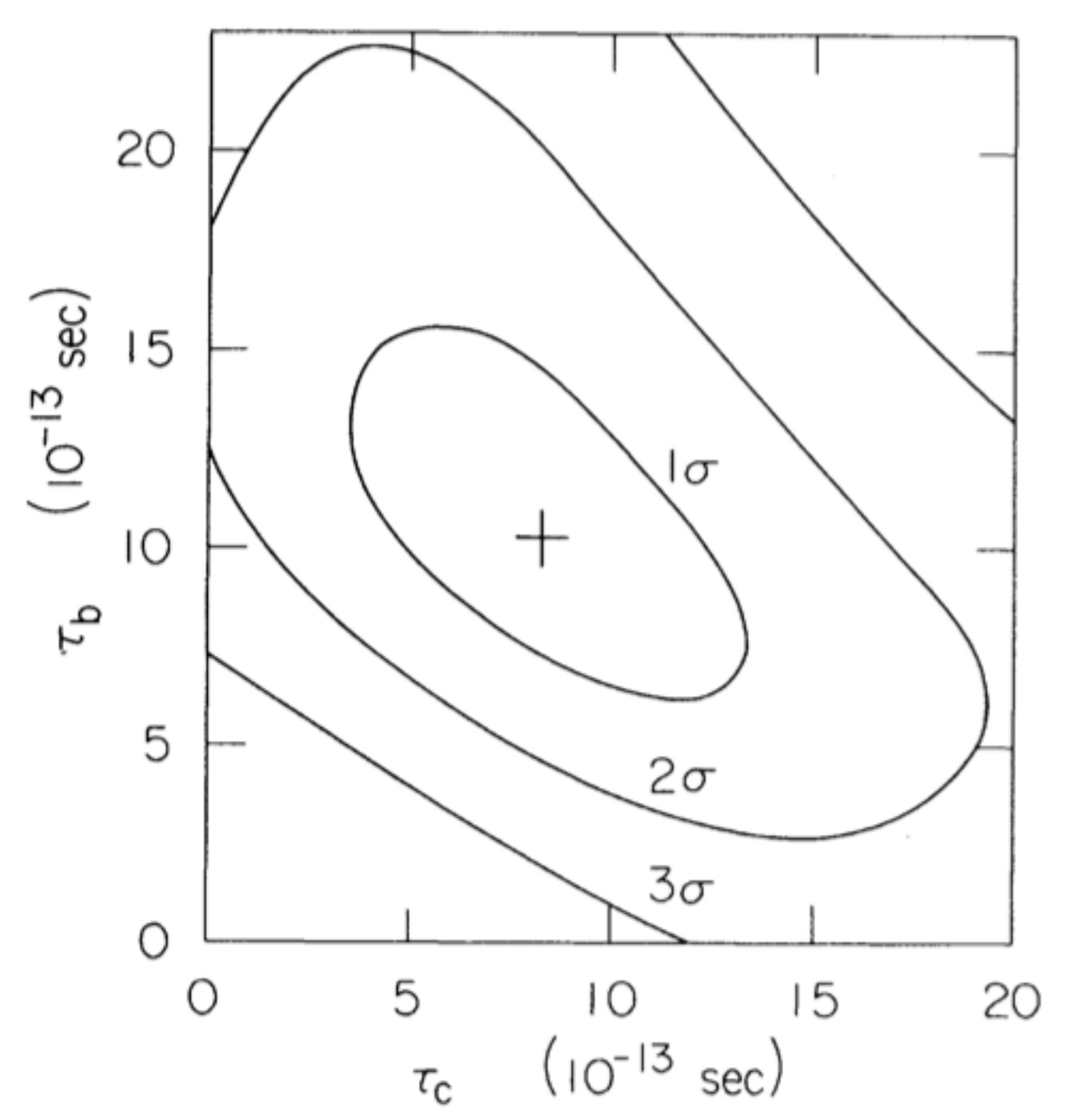}}
\end{minipage}
\begin{minipage}[b]{0.29\textwidth}
\begin{center}
Beauty lifetimes [fs]\\
$B^+: 1638 \pm 4$\\
$B^0: 1519 \pm 4$\\
$B_s: 1510 \pm 4$\\
$\Lambda_b: 1471 \pm 9$\\
$\Xi_b^-: 1572 \pm 40$\\
$\Xi_b^0: 1480 \pm 30$\\
$\Omega_b: 1640^{+180}_{-170}$

\vspace*{60pt}
\end{center}

\end{minipage}
\caption{Contours of equal likelihood for charm and beauty lifetimes from  the Mark~II Experiment~\cite{Lockyer:1983ev} are shown together with the 2019 world-average lifetimes for hadrons containing $c$ or $b$ quarks~\cite{Tanabashi:2018oca}.\label{fig:lifetimes}}
\end{figure}

In the same year, the MAC~\cite{Fernandez:1983az} and Mark II~\cite{Lockyer:1983ev} experiments operating at the PEP $e^+e^-$ collider at SLAC established an unexpectedly long $b$-hadron lifetime (see Figure~\ref{fig:lifetimes}), which implied a small value for the quark-mixing matrix element $\abs{V_{cb}} \approx 0.05$.

\begin{figure}
\centerline{\includegraphics[height=0.28\textheight]{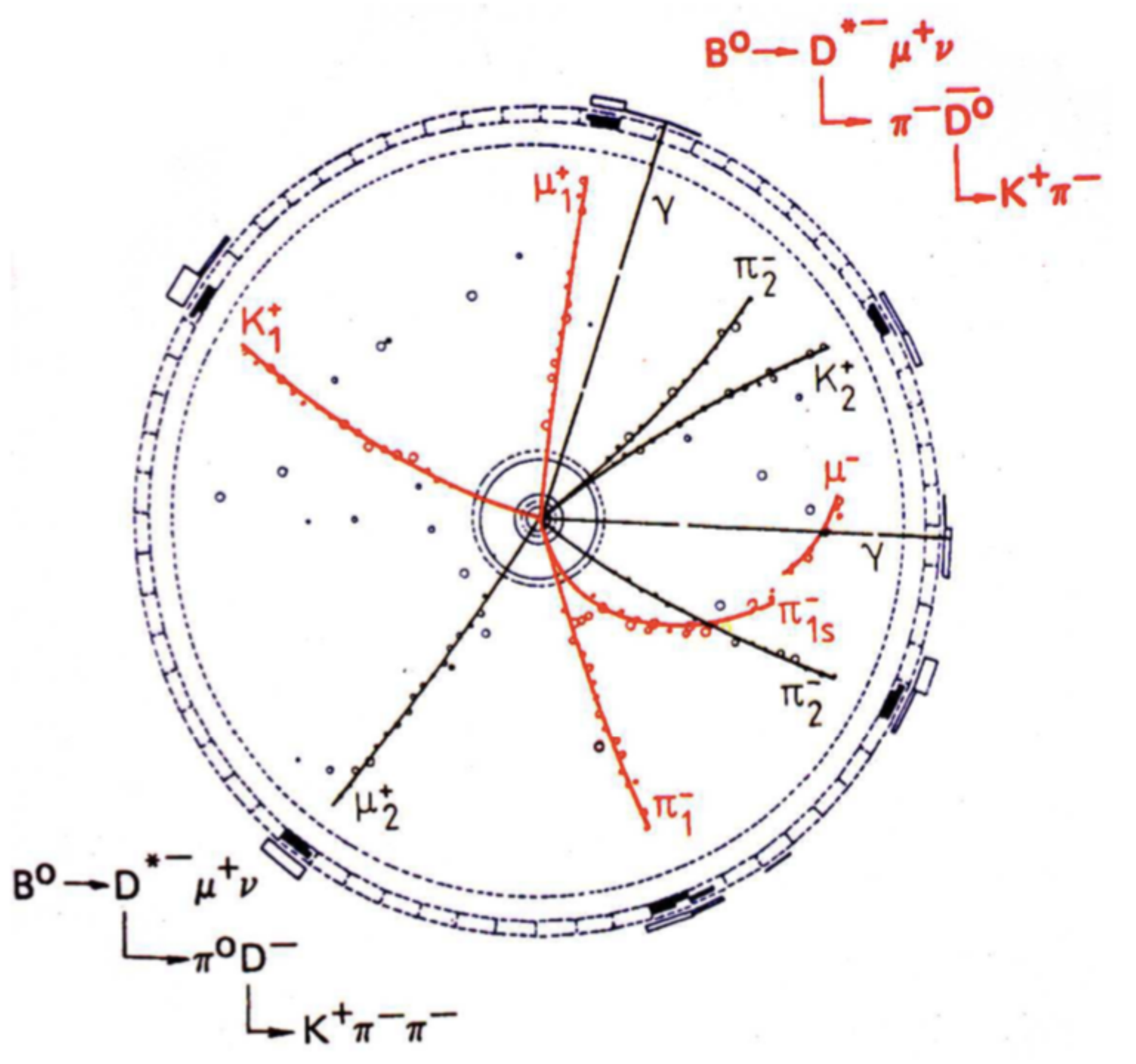}}
\caption{An explicitly mixed $\Upsilon(4S) \to B^0 B^0$ event, with fully reconstructed $B^0$-mesons, from Ref.~\cite{Prentice:1987ap}. \label{fig:golden}}
\end{figure}

In 1987, the UA1 experiment reported an excess of same-sign dimuons~\cite{Albajar:1986it} in $\sqrt{s}=630\gev$ collisions at the S$\bar{p}p$S Collider at CERN. The excess could be taken as evidence for $B^0\hbox{-}\bar{B}^0$  or $B_s^0\hbox{-}\bar{B}_s^0$ oscillations.
A golden event (Figure~\ref{fig:golden}) recorded in the ARGUS detector operating at DESY's DORIS II storage ring~\cite{Prentice:1987ap} unambiguously demonstrated $B^0\hbox{-}\bar{B}^0$ oscillations. 
Statistical evidence from like-sign dilepton events and events containing one reconstructed $B^0$ and an additional fast charged lepton indicated that $B^0$ mixing was substantial.

\begin{figure}
\centerline{\href{https://doi.org/10.1103/PhysRevD.45.3262}{\begin{overpic}[height=0.25\textheight]{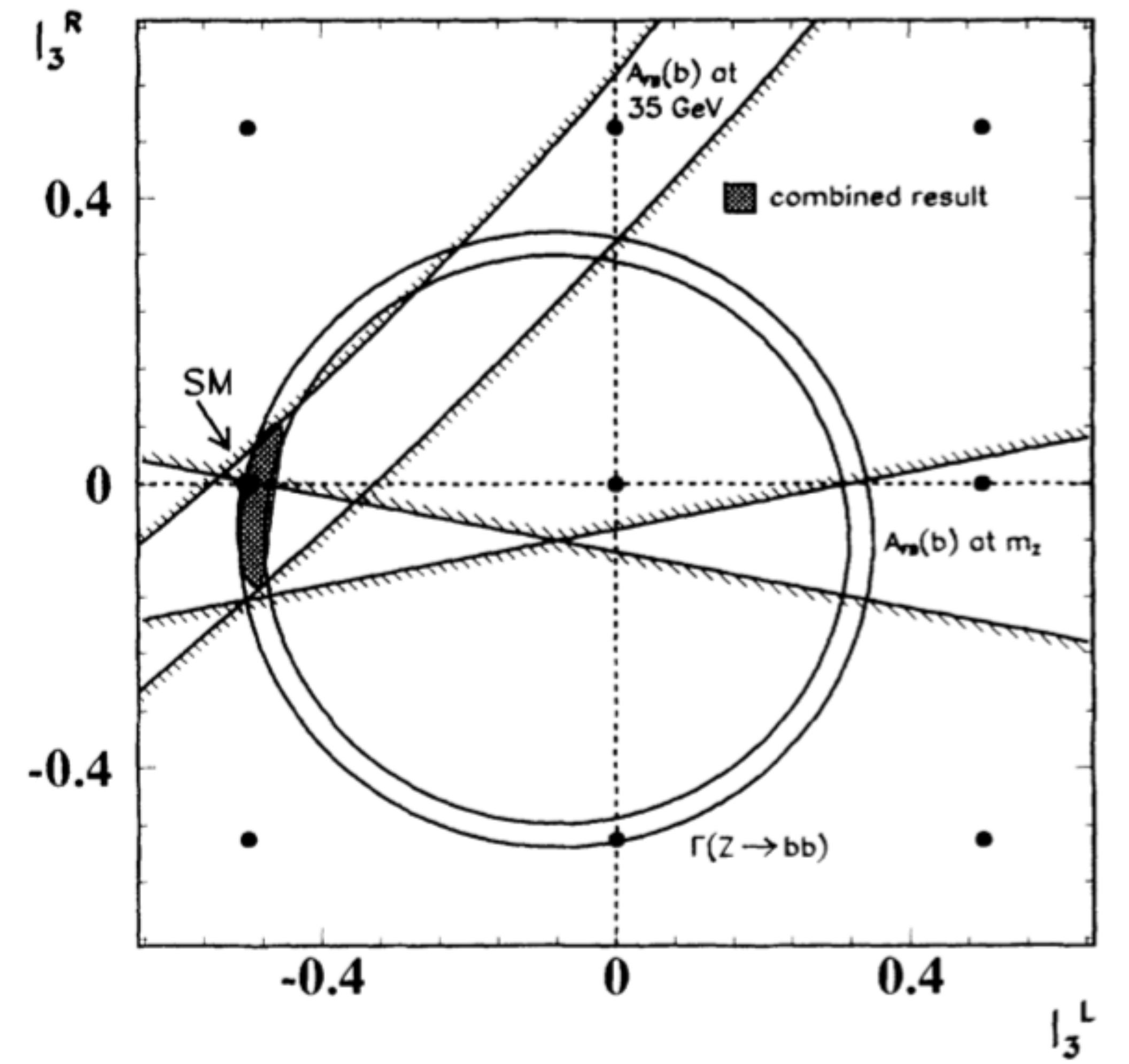}
\end{overpic}}}
\caption{Determination of the weak isospin of the $b$ quark (after Ref.~\cite{Schaile:1991tm}). \label{fig:weakiso}}
\end{figure}

Once we had established that the bottom quark carries charge $Q_b = -\cfrac{1}{3}$, it was natural to presume that it formed a weak-isospin doublet with an as yet undiscovered top quark. As experimental information accumulated about the neutral-current couplings of $b$, it became possible to back up that presumption~\cite{Schaile:1991tm}.  Working within a $V-A$ framework, we may generalize the left- and right-handed chiral couplings to 
$L_b \equiv 2I_{3\mathsf{L}} - 2Q_b\xw$, $R_b \equiv 2I_{3\mathsf{R}} - 2Q_b\xw$, where $\xw$ is the weak mixing parameter. Then we may deduce constraints on the chiral couplings, using these relations:
The partial width $\Gamma(Z^0 \to b\bar{b})$ measures $(L_b^2 + R_b^2)$, the forward--backward asymmetry $A_{b\bar{b}}^{\mathrm{(peak)}}$ in the reaction $e^+e^- \to b\bar{b}$ on the $Z^0$ peak is sensitive to $(L_b^2 - R_b^2)/(L_b^2 + R_b^2)$, and the forward-backward asymmetry in the $\gamma\hbox{-}Z$ interference regime  $A_{b\bar{b}}$ is proportional to $(R_b - L_b)$. A graphical representation of these constraints is given in Figure~\ref{fig:weakiso}. 
The three constraints overlap in a small region consistent with the standard-model expectation $(I_{3\mathsf{L}} = -\half, I_{3\mathsf{R}}=0)$, indicating that the $b$-quark is indeed the lower member of a weak-isospin doublet.

Expectations of appreciable \textsf{CP} violation in $B^0$ decays were fulfilled in 2001 in reports from BABAR~\cite{Aubert:2001nu} ($\sin{2\beta} \approx 0.59$) and Belle~\cite{Abe:2001xe} ($\sin{2\phi_1} \equiv \sin{2\beta} \approx 0.99$). \textsf{CP}-violating asymmetries have now been observed in dozens of decay modes, and the mixing parameter has settled to $\sin{2\beta}  =0.699 \pm 0.017$~\cite{Amhis:2019ckw}.
In 2006, the CDF Collaboration reported the definitive observation of  time-dependent $B_s^0\hbox{-}\bar{B}_s^0$ mixing, fixing the (very rapid) oscillation frequency at $\Delta m_{B^0_s}=17.77 \pm 0.10 \pm 0.07\,\hbar\!\ps^{-1}$~\cite{Abulencia:2006ze}, within one standard deviation of the current world average. Five years later, CDF reported the first detection of \textsf{CP} violation in charmless decays of $B_s$~\cite{Aaltonen:2011qt}. Thanks to decades of experimentation---in dialogue with theory---we now have a library of highly constraining precision tests of the Cabibbo--Kobayashi--Maskawa quark-mixing paradigm~\cite{fitters}, including the influence of penguin diagrams~\cite{Shifman:1975tn}.

\section{Mesons with Beauty and Charm}
\begin{figure}
\centerline{\includegraphics[height=0.20\textheight]{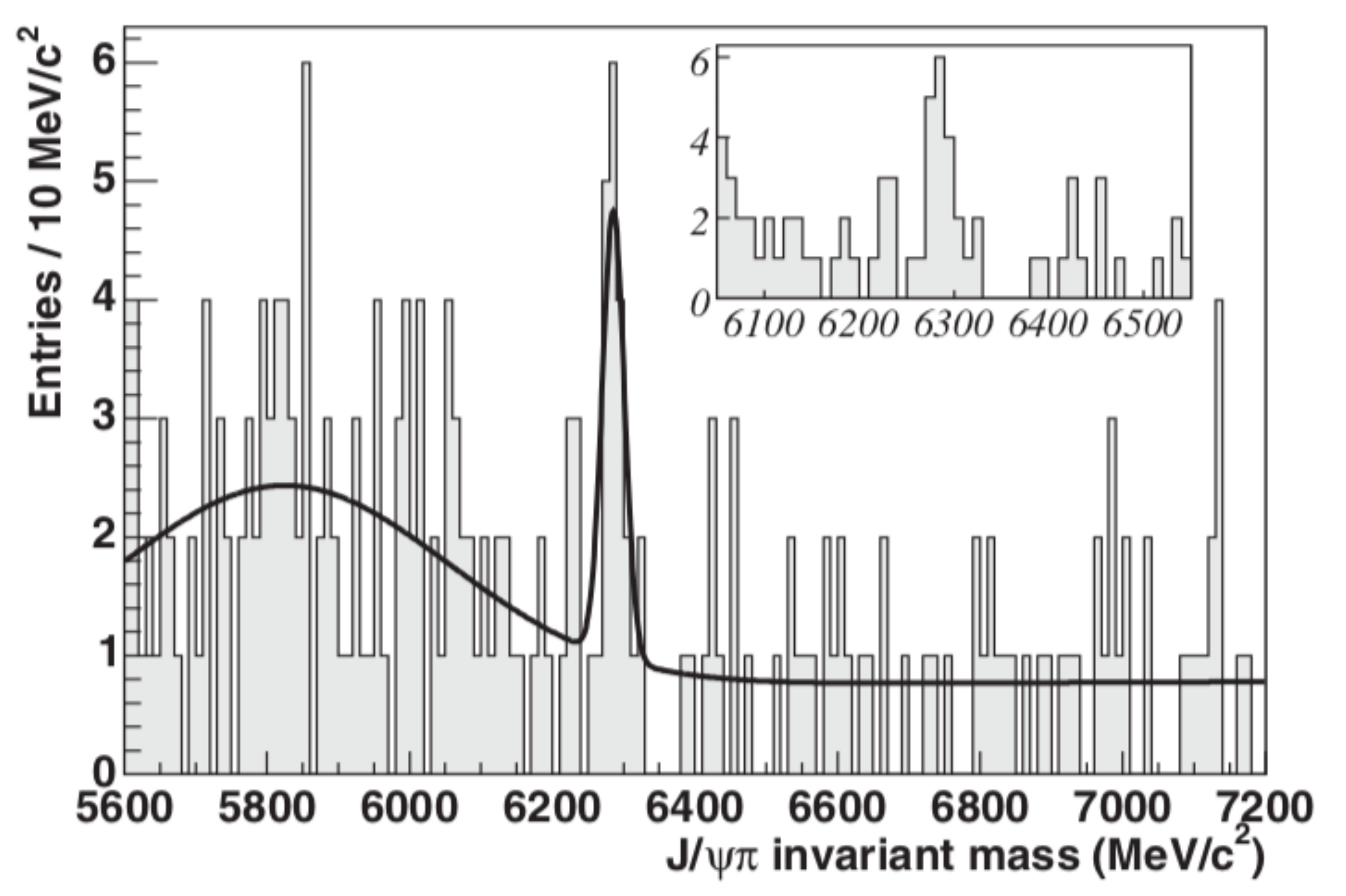}}
\caption{Invariant mass distribution of the $\jpsi\pi^\pm$ candidates.  The broad enhancement below $6.2\gev$ is attributable to partially reconstructed $B_c$ mesons (from Ref.~\cite{Abulencia:2005usa}).\label{fig:Bcdisc}}
\end{figure}

The $B_c$ family of mesons has attracted theoretical interest for decades because of its intermediate position between the \jpsi\ and $\Upsilon$ families, and because the unequal $c$ and $b$ quark masses make it sensitive to relativistic effects~\cite{Eichten:1994gt}. Since $B_c$ is composed of two different quarks, the annihilations into two- or three-gluon intermediate states that characterize $(c\bar{c})$ and $(b\bar{b})$ states cannot occur, so all excited states cascade to the ground state. The $B_c$ itself decays only through weak transitions $b\to c,\; c\to s,\hbox{ and }  b\bar{c} \to W^-$. It took a decade after the early theoretical studies until the CDF Collaboration was able to announce the reconstruction of $B_c \to \jpsi\pi^\pm$, shown in Figure~\ref{fig:Bcdisc}~\cite{Abulencia:2005usa}.
The long wait meant that the reconstructed mass, $M(B_c) = 6274.9\pm0.8\mev$, tested a precise prediction from lattice QCD~\cite{Allison:2004be}.

Until recently, the only evidence reported for a $(c\bar{b})$ excited state was presented by the ATLAS Collaboration~\cite{Aad:2014laa} in $pp$ collisions at $7\hbox{ and }8\tev$, in samples of $4.9\hbox{ and }19.2\fb^{-1}$. They observed a new state at $6842 \pm 7\mev$ in the $M(B_c^{\pm}\pi^{+}\pi^{-})-M(B_c^{\pm})-2M(\pi^{\pm})$ mass difference, with $B_c^{\pm}$ detected in the 
$J/\psi\, \pi^{\pm}$ mode. It was plausible to guess that  ATLAS might have observed the transition $B_c^*(2S) \to B_c^*(1S)\pi^+\pi^-$, missing the low-energy photon from the subsequent $B_c^* \to B_c \gamma$ decay, and that the signal is an unresolved combination of \spec{2}{3}{S}{1}\ and \spec{2}{1}{S}{0}\ peaks. A search by the LHC$b$ collaboration in $2\fb^{-1}$ of 8-TeV $pp$ data yielded no evidence for either $B_c(2S)$ state~\cite{Aaij:2017lpg}. 

\begin{figure}
\centerline{\raisebox{6pt}{\fcolorbox{white}{white}{\begin{overpic}[height=0.275\textheight]{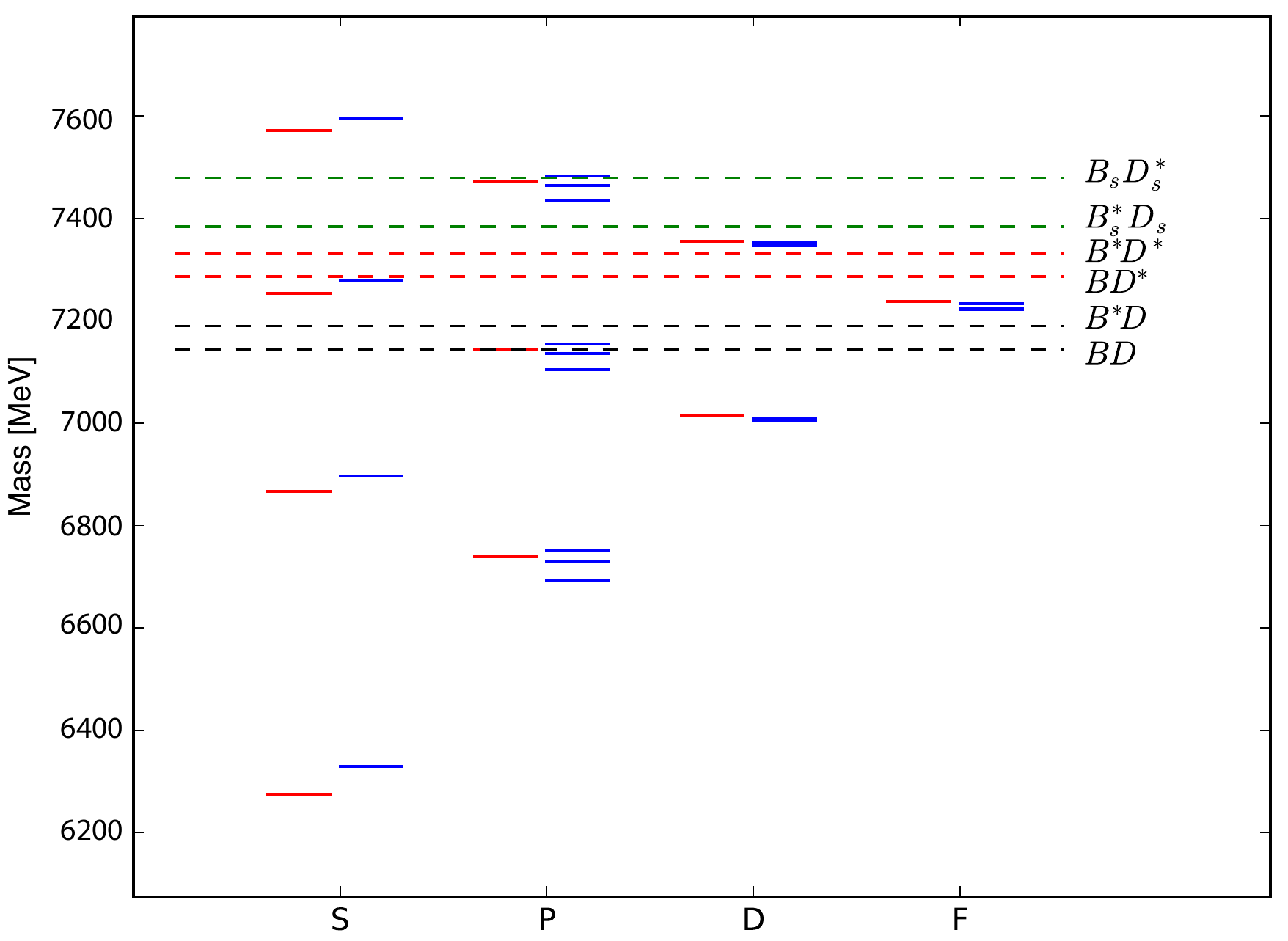}\put(40,65){{\small \textsf{$B_c$ spectrum}}}\end{overpic}}}}
\caption{Calculated $c\bar{b}$ spectrum, with (spin-singlet, spin-triplet) states  on the (left [red], right [blue]) for each orbital-angular-momentum family; dashed lines mark two-body open-flavor thresholds (Ref.~\cite{Eichten:2019gig}). \label{fig:bcspec}}
\end{figure}

The unsettled experimental situation and the advent of large new data sets motivated us to take a new look at the prospects for filling in the $B_c$ spectrum~\cite{Eichten:2019gig}. The spectrum shown in Figure~\ref{fig:bcspec}, calculated in a nonrelativistic potential model, indicates that 12--15 narrow $c\bar{b}$ levels might be found.
We calculated hadronic and electromagnetic cascade decay rates for the expected narrow states, and computed differential and integrated cross sections for the narrow $B_c$ levels in $pp$ collisions at the LHC.   Putting all these elements together, we discussed how to unravel the \spec{2}{}{S}{}\ levels and explored how higher levels might be observed.  

The CMS Collaboration~\cite{Sirunyan:2019osb} beat us to the arXiv, followed not long after by LHC$b$~\cite{Aaij:2019ldo}. These experiments published striking evidence for both $B_c(2S)$ levels, in the form of well-separated peaks in the $B_c\pi^+\pi^-$ invariant mass distribution, closely matching the theoretical template.
\begin{figure}
\centerline{{\includegraphics[width=0.45\textwidth]{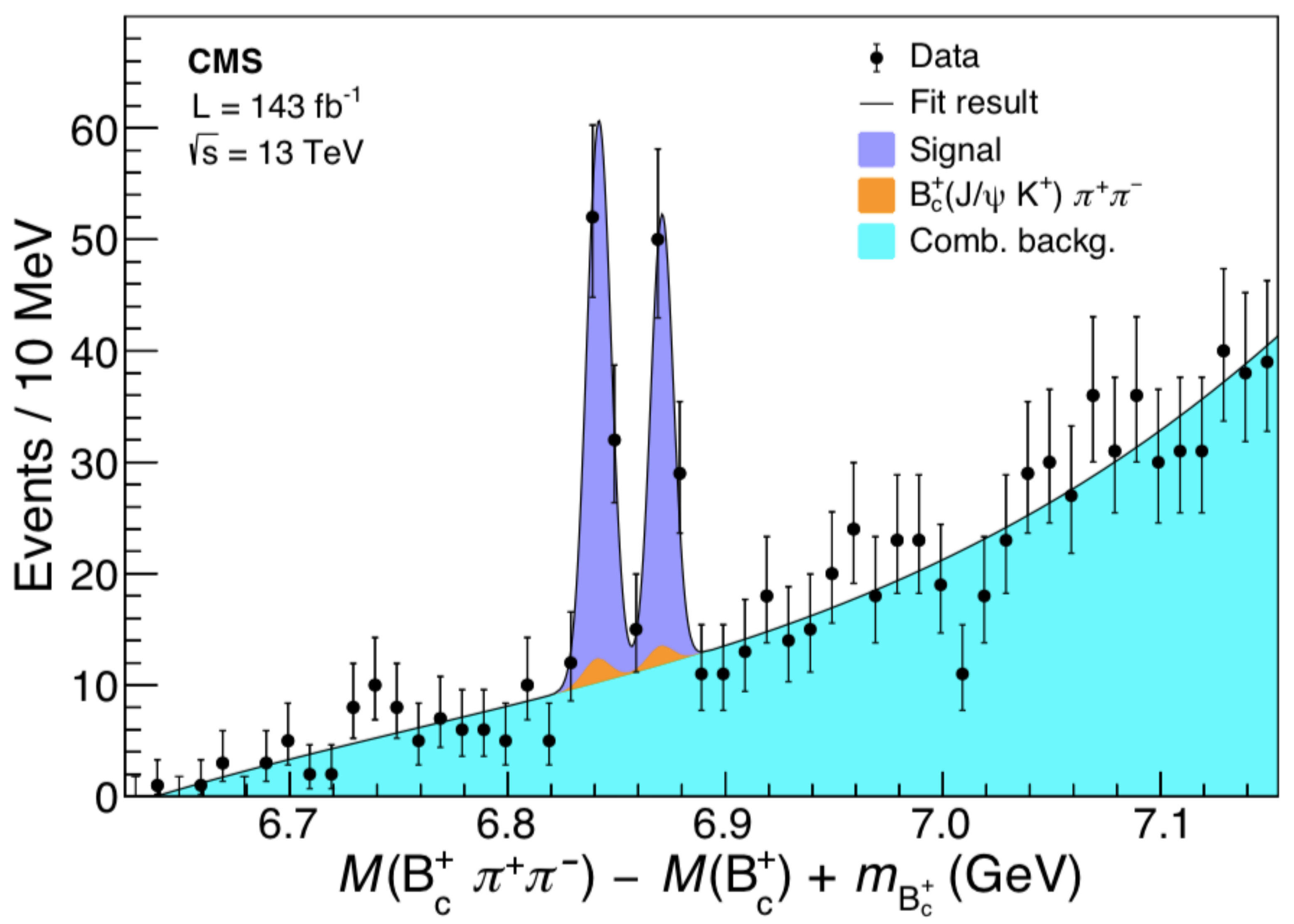}}\quad{\includegraphics[width=0.49\textwidth]{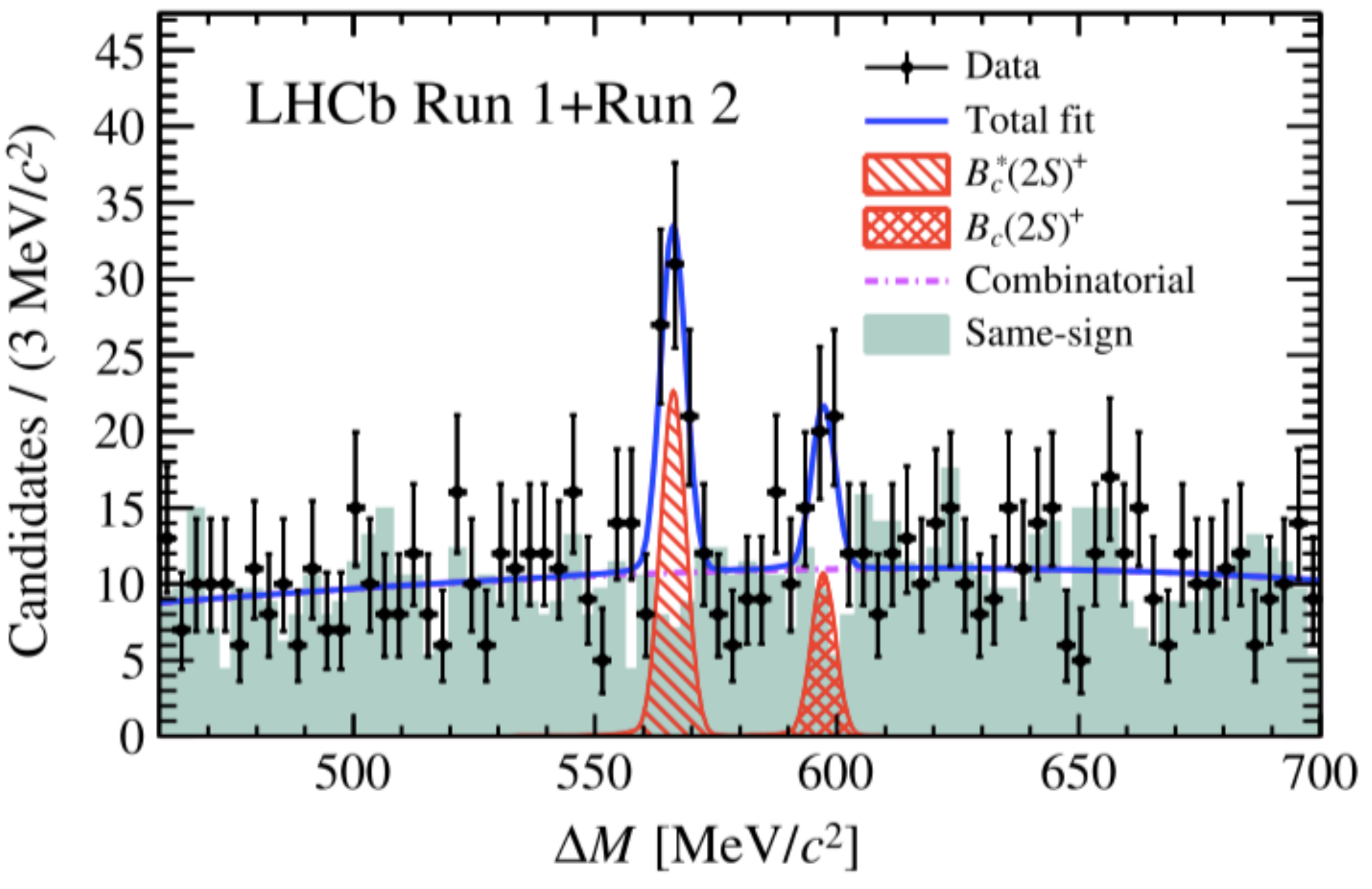}}}
\caption{(Shifted) $B_c \pi^+\pi^-$ invariant mass distributions reported by the CMS~\cite{Sirunyan:2019osb} (left panel) and LHC$b$~\cite{Aaij:2019ldo} (right panel) Collaborations. The (lower, upper) peak is interpreted as ($B_c^*(2S)$ , $B_c(2S)$) since  the photon energy in the $B_c^*(1S) \to B_c(1S) + \gamma$ transition goes undetected.  \label{fig:Bc2S}}
\end{figure}
The  difference of mass differences, $\Delta_{21}\equiv[M(B_c^{*\prime}) - M(B_c^{\prime})] - [M(B_c^{*}) - M(B_c)]$, sets the splitting between the peaks. We estimate $-23\mev$, whereas CMS measures $-29\mev$ and LHC$b$, $-31\mev$. The estimate depends on the not-yet-measured hyperfine splitting of the ground state,  $[M(B_c^{*}) - M(B_c)]$, for which we assumed $54\mev$, the consensus value of modern lattice QCD calculations. We look forward to further experimental analysis to determine the relative weights of the two peaks, which can test calculations of production and decay rates, and to studies of the $\pi^+\pi^-$ invariant mass distribution as next steps in $B_c$ spectroscopy.

Our calculations indicate that the \spec{3}{}{S}{} levels will lie above flavor threshold. The \spec{3}{1}{S}{0} state can decay into the final state $B^*\!D$  and the  \spec{3}{3}{S}{1} level has decays into both the $BD$ and $B^*\!D$ final states. However, it is conceivable that coupled-channel effects might push one or both states lower in mass. It is therefore worth examining the $B_c\pi^+\pi^-$ mass spectrum up through $7200\mev$ for indications of $\spec{3}{1}{S}{0} \to B_c\pi^+\pi^-$ and $\spec{3}{3}{S}{1} \to B_c^*\pi^+\pi^-$ lines. 

The \spec{3}{3}{P}{2} $(c\bar{b})$ state might be observed as a very narrow ($d$-wave) $BD$ line near open-flavor threshold, in the spirit of the LHC$b$ candidate~\cite{Aaij:2019evc} for the charmonium level $\spec{3}{3}{D}{3} \equiv \psi_3(3D) \to D\bar{D}$ at mass $M = 3842.72\pm0.16\pm0.12\mev$ and natural width $\Gamma = 2.79\pm0.51\pm0.35\mev$. We anticipate that one more narrow charmonium state, \spec{4}{3}{F}{4}, is to be found above flavor threshold, perhaps near $4054\mev$~\cite{Eichten:2004uh}.

It may in time become possible for experiments to detect some of the more energetic E1-transition photons that appear in $(c\bar{b})$ cascades. A particularly attractive target for experiment is the $\spec{2}{3}{P}{2}(6750) \to B_c^*\gamma$ line, because of the favorable production cross section, branching fraction, and 409-MeV photon energy. The $\spec{3}{3}{P}{2}(7154) \to B_c^*\gamma(777\mev)$ line is favored for its  high photon energy, which may be a decisive advantage for detection.
\section{Tetraquarks stable against strong decays}
The proposition that stable or nearly stable multiquark states containing heavy flavors might exist goes back nearly four decades~\cite{Ader:1981db}. The discovery of the doubly-charmed baryon $\Xi_{cc}^{++}(3621)$ in the LHC$b$ experiment~\cite{Aaij:2017ueg} has provoked a new wave of interest in exotic mesons containing two heavy quarks.
Estia Eichten and I have examined the possibility of unconventional tetraquark configurations for which all strong decays are kinematically forbidden~\cite{Eichten:2017ffp}. In the heavy-quark limit, stable---hence exceedingly narrow---$Q_iQ_j \bar q_k \bar q_l$ mesons must exist. To apply this insight, we take into account  corrections for finite heavy-quark masses to deduce which tetraquark states containing $b$ or $c$ quarks might be stable. The most promising candidate is a $J^P=1^+$ isoscalar double-$b$ meson, $\mathcal{T}^{\{bb\}-}_{[\bar u \bar d]}$. 

\begin{figure}
\centerline{\includegraphics[height=0.125\textheight]{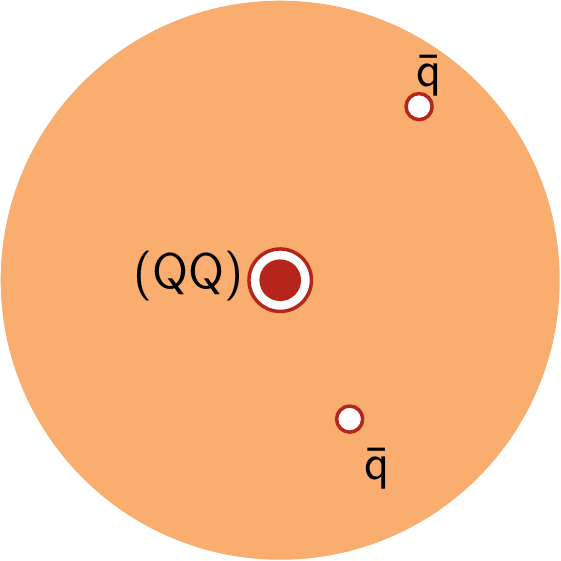}\qquad{\includegraphics[height=0.125\textheight]{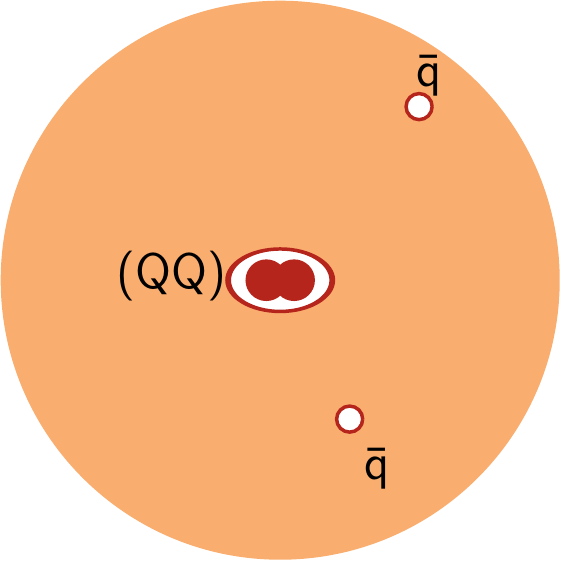}}\qquad{\includegraphics[height=0.125\textheight]{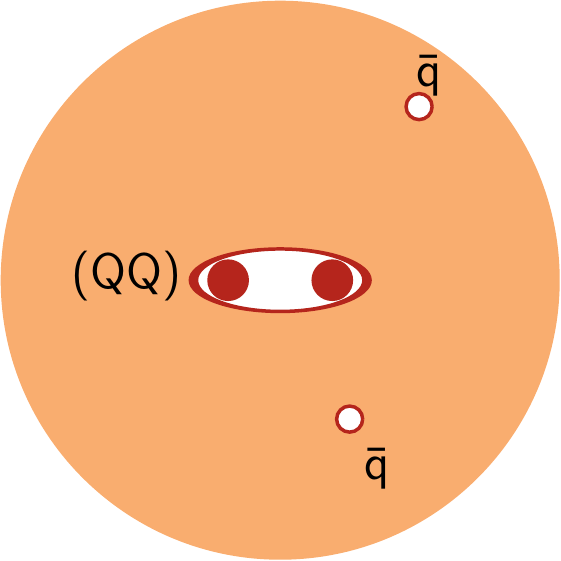}}\qquad{\raisebox{0pt}{\includegraphics[height=0.125\textheight]{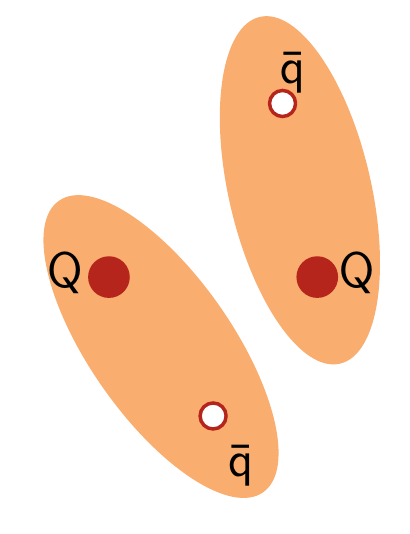}}}}
\caption{Schematic evolution of  a $Q_iQ_j \bar q_k \bar q_l$ state as the heavy-quark masses decrease (and the mean separation between the heavy quarks increases) from left to right.\label{fig:dhtq}}
\end{figure}

Let us start with the observation that one-gluon-exchange between a pair of color-triplet heavy quarks is attractive for $(QQ)$ in a color-$\mathbf{\bar{3}}$ configuration and repulsive for the color-$\mathbf{6}$ configuration. The strength of the $\mathbf{\bar{3}}$ attraction is half that of the corresponding $(Q\bar{Q})$ in a color-$\mathbf{1}$. This means that in the limit of very heavy quarks, we may idealize the color-antitriplet $(QQ)$ diquark as a stationary, structureless color charge, as depicted in the leftmost panel in Figure~\ref{fig:dhtq}.
In that case, we can separate the strong dynamics binding the diquark from the long-range color interaction by which the light antiquarks interact with each other and are bound to the diquark ``nucleus.''
For sufficiently heavy quarks $Q$, a $Q_iQ_j \bar q_k \bar q_l$ tetraquark meson is stable against strong decays, as we can show by considering possible decay modes. First, we note that dissociation into two heavy--light mesons is kinematically forbidden. The $\mathcal{Q}$ value for the decay is \break
$\mathcal{Q} \equiv m(Q_i Q_j \bar q_k \bar q_l) - [m(Q_i \bar q_k) + m(Q_j \bar q_l)] = 
\Delta(q_k, q_l) - \half\!\left(\cfrac{2}{3}\alphas\right)^2\![1 + O(v^2)]\overline M + O(1/\overline M)$,
where $\Delta(q_k, q_l)$, the contribution due to light dynamics, becomes independent of the heavy-quark masses,  $\overline M \equiv (1/{m_Q}_i + 1/{m_Q}_j)^{-1}$ is the reduced mass of $Q_i$ and $Q_j$, and \alphas\ is the strong coupling.  For large enough values of $\overline M$, the middle term on the right-hand side dominates, so the tetraquark is stable against decay into two heavy-light mesons.

What of the other possible decay channel, a doubly heavy baryon plus a light antibaryon,
$(Q_iQ_j \bar q_k \bar q_l) \to (Q_iQ_j q_m) + (\bar q_k \bar q_l\bar q_m)$? For very heavy quarks, the contributions of $Q$ motion and spin to the tetraquark mass are negligible.
Since the $(QQ)$ diquark is a color-antitriplet, heavy-quark symmetry tells us that 
\begin{equation}
m(Q_iQ_j \bar q_k \bar q_l) - m(Q_iQ_j q_m) = m(Q_x q_k q_l) - m(Q_x \bar q_m).
\label{eq:hqsrel}
\end{equation}
Taking into account finite-mass corrections prescribed by heavy-quark symmetry, we can use measured masses to show that the right-hand side is in every case smaller than the mass of the lightest antibaryon, $\bar p$, so no decay to a doubly heavy baryon and a light antibaryon is kinematically allowed. \emph{With no open channels in the heavy-quark limit, stable $Q_iQ_j \bar q_k \bar q_l$ mesons must exist.} 

To assess the implications for the real world, we must first test whether it makes sense to idealize the $(QQ)$ diquark as a tiny, structureless, color-antitriplet color source. As the separation between the heavy quarks increases, the light-antiquark cloud screens the $Q_iQ_j$ interaction, altering the $\mathbf{\bar{3}, 6}$ mix, and eventually leading to the fission of the $(Q_iQ_j \bar q_k \bar q_l)$ state into a pair of heavy--light mesons. These changes are indicated in the progression from left to right in Figure~\ref{fig:dhtq}. Using a half-strength Coulomb$+$linear quarkonium potential, we verified that the rms core radii are small on the expected tetraquark scale: $\langle r^2\rangle^{1/2} = 0.19\fm\, (bb);\; 0.24\fm\, (bc); \;0.28\fm\, (cc)$. This assessment is consistent with other results~\cite{Czarnecki:2017vco}.

If we had access to measured masses for all of the doubly heavy baryons, Eqn.~(\ref{eq:hqsrel}), with controlled finite-mass corrections, would yield predictions of tetraquark masses, without the intervention of any models. That is not yet the case; only one example, $\Xi_{cc}^{++}$, can be considered established. While waiting for experiments to provide more comprehensive information, we use as inputs the model calculations of doubly-heavy baryon masses by Karliner \& Rosner~\cite{Karliner:2014gca}.

We find two real-world candidates for stable tetraquarks: the axial vector $\{bb\}[\bar u \bar d]$  meson, $\mathcal{T}^{\{bb\}-}_{[\bar u \bar d]}$ bound by $121\mev$ against strong decays, and the axial vector $\{bb\}[\bar u \bar s]$ and $\{bb\}[\bar d \bar s]$ mesons bound by $48\mev$. 
Given the provisional doubly heavy baryon masses, we expect all the other $Q_iQ_j \bar q_k \bar q_l$ tetraquarks to lie at least $78\mev$  above the corresponding thresholds for strong decay.  Promising final states in which to search for stable tetraquarks  include $\mathcal{T}^{\{bb\}}_{[\bar u \bar d]}(10482)^-\! \to \Xi^0_{bc}\bar{p}$, $B^-D^+\pi^-$, and $B^-D^+\ell^-\bar{\nu}$ (which establishes a weak decay), $\mathcal{T}^{\{bb\}}_{[\bar u \bar s]}(10643)^-\! \to \Xi^0_{bc}\bar{\Sigma}^-$,  $\mathcal{T}^{\{bb\}}_{[\bar d \bar s]}(10643)^0\! \to \Xi^0_{bc}(\bar{\Lambda},\bar{\Sigma}^0)$, and so on. Observing a weakly decaying double-beauty state would establish the existence of tetraquarks and illuminate the role of heavy color-$\mathbf{\bar{3}}$ diquarks as hadron constituents.

Heavier $bb \bar q_k \bar q_l$ states, as well as $bc \bar q_k \bar q_l$ and  $cc \bar q_k \bar q_l$  might be seen as ``wrong-sign''  $Q\bar{q} + Q\bar{q}$ resonances in double-flavor $DD, DB, BB$ combinations near threshold. (In their model calculations, Karliner and Rosner~\cite{Karliner:2017qjm} estimate somewhat deeper binding, and so point to additional $bc$ and $cc$ candidates.) For example, the double-charge, double charm $J^P = 1^+\; \mathcal{T}^{\{cc\}++}_{[\bar d \bar s]}(4156) \!\to D^+ D_s^{*+}$ would constitute {prima facie evidence} for a non-$q\bar{q}$ level. Other promising cases include the $1^+\; \mathcal{T}^{\{bb\}}_{\{\bar q_k \bar q_l\}}(10681)^{0,-,--}$, $1^+\; \mathcal{T}^{\{bc\}}_{[\bar u \bar d]}(7272)^0$,  
$0^+\; \mathcal{T}^{[bc]}_{[\bar u \bar d]}(7229)^0$,  and $1^+\; \mathcal{T}^{\{cc\}}_{[\bar u \bar d]}(3978)^+$.
None of this will be easy, but both experiment and theory have much to do along the way.

\renewcommand\theenumi{\small $\mathcal{T}$\arabic{enumi}}%
\begin{enumerate}[noitemsep]

\item Look for double-flavor resonances near threshold.

\item Measure cross sections for final states containing 4 heavies: $Q_i\bar{Q}_iQ_j\bar{Q}_j$.

\item Discover and determine masses of doubly-heavy baryons.
We need this information  to implement the heavy-quark-symmetry calculation of tetraquark masses embodied in Eqn.~(\ref{eq:hqsrel}). The doubly heavy baryons are also of intrinsic interest: in the heavy-quark limit, they resemble heavy--light mesons, with the added possibility of $(QQ)$ core excitations. 

\item Resolve the uncertainty surrounding $\Xi_{cc}^+$  between LHC$b$~\cite{Aaij:2019jfq} and SELEX~\cite{Mattson:2002vu}.

\item Find stable tetraquarks through  weak decays. A rough guess for the lifetimes is $\sim 1\ps$. 

\item Develop expectations for production (cf. Ref.~\cite{Ali:2018ifm}). 

\item Refine lifetime estimates for stable states.

\item Understand how color configurations evolve with $QQ$ (and $\bar{q}\bar{q}$) masses. 

\item Does it make sense to consider body plans such as {$(Q_iQ_j)(Q_kQ_l)(Q_mQ_n)$} in the heavy-quark limit? This is an example of a six-quark state with baryon number $B=2$, but a $Q_pQ_qQ_r$ color structure, if the three-diquark configuration should dominate.

\end{enumerate}

\section{Flavor: the problem of identity}
The essence of the problem is easy to state, but not so easy to answer: What makes an electron an electron, a top quark a top quark, a neutrino a neutrino?
{We do not have a clear view of how to approach the diverse character of the constituents of matter.} The CKM paradigm is an extraordinarily fruitful framework in the hadron sector, but there are many parameters. We have no clue what determines them, nor at what energy scale they are  set.
Even if the Higgs mechanism explains \emph{how}  masses and mixing angles arise, we do not know \emph{why} they have the values we observe. In this sense, all fermion masses---beginning with the electron mass---cry out for physics beyond the standard model!

Our ability to calculate within the standard model rests on knowing the values of many parameters: 3 Coupling parameters, $\alphas$, $\alpha_{\mathrm{em}}$, $\sin^2\theta_{\mathrm{W}}$;  2 Parameters of the Higgs potential; 1 Vacuum phase (QCD); 6 Quark masses, 3  Quark mixing angles, 1  CP-violating phase; 3  Charged-lepton masses, 3  Neutrino masses, 3  Leptonic mixing angles, 1  Leptonic CP-violating phase (probably plus two Majorana phases), for a total of $26^+$  seemingly arbitrary parameters. Of these, twenty concern the flavor sector. The problem of identity is rich in questions:
\renewcommand\theenumi{\small F\arabic{enumi}}%
\begin{enumerate}[noitemsep]
\item Can we find evidence of right-handed charged-current interactions?   Is nature built on a fundamentally asymmetrical plan, or are the right-handed weak interactions simply too feeble for us to have observed until now, reflecting an underlying hidden symmetry? 
\item What is the relationship of left-handed and right-handed fermions?
\item Are there additional electroweak gauge bosons, beyond $W^\pm$ and $Z$?
\item Are there additional kinds of matter? 
\item Is charged-current universality exact?   What about lepton-flavor universality?
\item Where are flavor-changing neutral currents in quark transitions? In the standard model, these are absent at tree level and highly suppressed by the Glashow--Iliopoulos--Maiani mechanism~\cite{Glashow:1970gm}. They arise generically in proposals for physics beyond the standard model, and need to be controlled. And yet we have made no sightings!   {Why not?}  The focus for now is on studies of $B_{s,d} \to \mu^+\mu^-$ and $K^+ \to \pi^+\nu\bar{\nu}$.

\item How well can we test the standard-model correlation among the quark-mixing matrix parameter $\gamma$, $\mathcal{B}(K^+ \to \pi^+\nu\bar{\nu}$), and $\mathcal{B}(B_s \to \mu^+\mu^-)$?
\item Have we found the ``periodic table'' of elementary particles?
\item What do generations mean? Is there a family symmetry?
\item Why are there three families of quarks and leptons? (Is it so?)
\item Are there new species of quarks and leptons, possibly carrying exotic charges?
\item Is there any link to a dark sector?
\item What will resolve the disparate values of $\abs{V_{ub}}$ and $\abs{V_{cb}}$ measured in inclusive and exclusive decays?
\item Is the $3\times 3$ (CKM) quark-mixing matrix unitary?
\item Why is isospin a good symmetry? What does it mean?
\item Can we find evidence for charged-lepton flavor violation in lepton decays?
\item Will we establish and diagnose a break in the standard model? 
\item {Do flavor parameters \emph{mean} anything at all?}   Contrast the landscape perspective.
\item If flavor parameters have meaning (beyond engineering information), {what is the meta-question?}

\end{enumerate}

\section{Top matters}
The top quark touches many topics in particle physics and presents us with many questions.
\renewcommand\theenumi{\small $t$\arabic{enumi}}%
\begin{enumerate}[noitemsep]
\item How stringently will refined measurements of $M_W$ and $m_t$ test the electroweak-theory prediction for  $M_H$?
\item How much can we tighten the $m_t$-$M_W$-$M_H$ constraints?
\item Does top's large $Ht\bar{t}$ (Yukawa) coupling imply a special role in electroweak symmetry breaking? How does it influence $t\bar{t}$ dynamics?
Does the large value of $m_t$ make top an outlier or the only normal fermion?
\item How well can we constrain $V_{tb}$ in single-top production, and elsewhere?
\item How much can we refine our knowledge of $V_{td}$ and $V_{ts}$?
\item How complete is our understanding of $t\bar{t}$ production in QCD: total and differential cross sections, charge asymmetry, spin correlations, etc.?
\item What might we learn from ``dead-cone'' studies using boosted tops~\cite{Dokshitzer:1991fd}
?
\item How well can we constrain the top-quark lifetime~\cite{ATLAS:2019onj}?  How {free} is $t$?
\item Are there (vestiges of) $t\bar{t}$ resonances?
\item Can we find evidence of flavor-changing top decays $t \to (Z, \gamma)(c,u)$? 
\end{enumerate}

\section{Electroweak symmetry breaking and the Higgs sector}
In a few short years, the LHC experiments have given us a wealth of information about the Higgs boson~\cite{HiggsInfo}. We can summarize by saying that the evidence is developing as it would if $H(125\gev)$ were the textbook scalar responsible for electroweak symmetry breaking and the generation of fermion masses and mixings. Here is
a list of questions about electroweak symmetry breaking and the Higgs sector that we must answer to approach a final verdict about how closely $H(125)$ matches the textbook Higgs boson~\cite{Dawson:2018dcd}.
\renewcommand\theenumi{\small $H$\arabic{enumi}}%
\begin{enumerate}[noitemsep]
\item Is $H(125)$ the only member of its clan? Might there be others---charged or neutral---at higher or lower masses?
\item Does $H(125)$ fully account for electroweak symmetry breaking? Does it match standard-model branching fractions to gauge bosons? Are absolute couplings to $W$ and $Z$ as expected in the standard model?
\item Are all production rates as expected?   {Any surprise sources of $H(125)$?}
\item What accounts for the immense range of fermion masses?
\item Is the Higgs field the only source of fermion masses?   Are fermion couplings proportional to fermion masses?    {$\mu^+\mu^-$ soon?}  How can we detect $H \to c\bar{c}$?  What about $H \to e^+e^-$, which would give new insight into the finiteness of the Bohr radius and the origins of valence bonding?
\item What role does the Higgs field play in generating neutrino masses? 
\item Can we establish or exclude decays to new particles? Does $H(125)$ act as a portal to hidden sectors?   {When can we measure $\Gamma_H$?}\phantom{M}
\item Can we detect flavor-violating decays ($\tau^\pm\mu^\mp$, \ldots)?
\item Do loop-induced decays ($gg, \gamma\gamma, \gamma Z$) occur at standard-model rates?
\item What can we learn from rare decays ($\jpsi\,\gamma, \Upsilon\,\gamma$, \ldots)?
\item Does the electroweak vacuum seem stable, or suggest a new physics scale?
\item Can we find signs of new strong dynamics or (partial) compositeness?
\item {Can we establish the $HHH$ trilinear self-coupling?}
\item {How well can we test the notion that $H$ regulates Higgs--Goldstone scattering, i.e., tames the high-energy behavior of $WW$ scattering?}
\item {Is the electroweak phase transition first-order?}
\end{enumerate}

\section{Future instruments}
Our experimental future demands both diversity and scale diversity, but an important driver of progress will be the next great accelerator. Every one of us should take time to explore the possibilities, form an opinion, and communicate it to our colleagues in particle physics and beyond. I have given an inventory of frontier machines in a recent essay~\cite{Quigg:2018llo} and series of seminars~\cite{PandQ}, where you may find pointers to detailed information. Now, the great question:
How do \emph{you} assess the scientific potential \emph{for Beauty and in general} of
 \begin{quote}{
(a) The High-Luminosity LHC?\\
(b) The High-Energy LHC?\\
(c) A 100-TeV $pp$ Collider (FCC-hh or SppC)?\\
(d) A 250-GeV ILC?\\
(e) A circular Higgs factory (FCC-ee or CEPC)?\\
(f) A 380-GeV CLIC?\\
(g) A $\mu^+\mu^- \to H$ Higgs factory?\\
(h) LHeC / FCC-eh? (or an electron--ion collider?)\\
(i) A muon-storage-ring neutrino factory?\\
(j) A multi-TeV muon collider?\\
(k) The instrument of your dreams?}
\end{quote}

\acknowledgments
I thank the Beauty2019 Organizing Committee, particularly Robert Fleischer and Guy Wilkinson, for their kind invitation to present the opening lecture. Congratulations to our Ljubljana hosts, led by Bostjan Golob, for their flawless preparations and peerless hospitality. I join all participants in looking forward to Beauty2020, where we anticipate the first wave of results from Belle II.

This work was supported by Fermi Research Alliance, LLC under Contract No. DE-AC02-07CH11359 with the U.S. Department of Energy, Office of Science, Office of High Energy Physics, and by the Munich Institute for Astro- and Particle Physics (MIAPP), which is funded by the Deutsche Forschungsgemeinschaft (DFG, German Research Foundation) under Germany's Excellence Strategy -- EXC-2094 -- 390783311. I am grateful to the Visiting Professor Program of the Bavarian State Ministry for Science, Research, and the Arts 
and also the Institute for Advanced Study at Technical University Munich where much of this talk was prepared. 
I thank Estia Eichten for a long and  enlightening collaboration.

\end{document}